\documentclass[USenglish,twocolumn]{article}
\usepackage[utf8]{inputenc}				%(only for the pdftex engine)
\usepackage[big,online]{dgruyter}	%values: small,big | online,print,work
\usepackage{lmodern}
\usepackage{microtype}
\usepackage[numbers,square,sort&compress]{natbib}
\usepackage{amsmath}
\usepackage{amsfonts}
\usepackage{graphicx}
\usepackage{epsfig}
\usepackage{color}
\usepackage{times}
\theoremstyle{dgthm}

\theoremstyle{dgdef}

\begin{document}

%%%--------------------------------------------%%%
  \articletype{Research Article}
  \received{Month	DD, YYYY}
  \revised{Month	DD, YYYY}
  \accepted{Month	DD, YYYY}
  \journalname{De~Gruyter~Journal}
  \journalyear{YYYY}
  \journalvolume{XX}
  \journalissue{X}
  \startpage{1}
  \aop
  \DOI{10.1515/sample-YYYY-XXXX}
%%%--------------------------------------------%%%

\title{Topological optomechanical amplifier with synthetic $\mathcal{PT}$-symmetry}
\runningtitle{ Topological optomechanical amplifier with synthetic $\mathcal{PT}$-symmetry\ \ }

\author[1]{Jian-Qi Zhang}%\ use * to mark the author as the corresponding author
\author[1,2,3]{Jing-Xin Liu}
\author[4]{Hui-Lai Zhang}
\author[5]{Zhi-Rui Gong}
\author[6]{Shuo Zhang}
\author[2]{Lei-Lei Yan}
\author*[2]{Shi-Lei Su}
\author*[4]{Hui Jing}
\author*[1]{Mang Feng}
\runningauthor{J.-Q. Zhang, J.-X. Liu, et al.}
\affil[1]{\protect\raggedright
State Key Laboratory of Magnetic Resonance and Atomic and
Molecular Physics, Wuhan Institute of Physics and Mathematics, Innovation Academy of Precision Measurement Science and Technology, Chinese
Academy of Sciences, Wuhan 430071, China, e-mail: mangfeng@wipm.ac.cn (M. Feng). The first two authors contributed equally to this work.}
\affil[2]{\protect\raggedright
School of Physics, Zhengzhou University,
Zhengzhou 450001, China, e-mail:slsu@zzu.edu.cn (S.-L. Su).}
\affil[3]{\protect\raggedright
National Laboratory of Solid State Microstructures, School of Physics, Nanjing University, Nanjing 210093, China.}
\affil[4]{Key Laboratory of Low-Dimensional Quantum Structures and Quantum Control of Ministry of Education, Department of Physics and Synergetic Innovation Center for Quantum Effects and Applications, Hunan Normal University, Changsha 410081, China, e-mail: jinghui73@foxmail.com (H. Jing)}
\affil[5]{\protect\raggedright
The College of Physics and Optoelectronic Engineering, Shenzhen University,
Shenzhen 518060, China}
\affil[6]{\protect\raggedright
Henan Key Laboratory of Quantum Information and Cryptography, Zhengzhou, 450001, China.}

\abstract{We propose how to achieve synthetic $\mathcal{PT}$ symmetry in optomechanics without using any active medium.
We find that harnessing the Stokes process in such a system can lead to the emergence of exceptional point (EP),
i.e., the coalescing of both the eigenvalues and the eigenvectors of the system.
 By encircling the EP, both non-reciprocal optical amplification and chiral mode switching can be achieved. As a result,
our synthetic $\mathcal{PT}$-symmetric optomechanics works as a topological optomechanical amplifier. This provides a surprisingly
simplified route to realize $\mathcal{PT}$-symmetric optomechanics, indicating that a wide range of EP devices can
be created and utilized for various applications such as topological optical engineering and nanomechanical processing or sensing.}


\maketitle

\section{Introduction}
Unconventional effects of exceptional points (EPs), i.e., non-Hermitian spectral degeneracies at which the eigenvalues and their eigenvectors coalesce, as revealed in recent years~\cite{science-363-42,nmater-18-783,nphys-14-11,prl-124-053901,ol-42-1556,nature-576-65,nature-576-70,nl-20-7594,science-372-72,prl-118-093002, prl-101-080402,nphys-6-192,nphys-10-394,nphoton-8-524,pra-88-053810,light-8-88,prl-103-093902,science-370-1077,prl-114-253601,
science-364-878,prl-123-213903,nature-526-554,prl-80-5243,rpp-70-947,prxq-2-020307,arxiv}, have radically changed our understanding of
complex systems and led to important applications. Novel EP devices have been fabricated and utilized for realizing, e.g., ultra-sensitive metrology~\cite{prl-117-110802,nature-548-192,prl-112-203901,nature-548-187},
single-mode lasing~\cite{science-346-975,science-346-972,prl-113-053604,nphoton-12-479},
loss-induced transparency~\cite{oe-19-25199,science-346-328}, and wireless power transfer~\cite{nature-546-387,prap-15-014009}. In particular, EP-enabled exotic topological effects have attracted intense interests~\cite{prx-9-041015,prb-99-235112,prl-121-026808,prl-121-086803,rmp-93-015005}, such as non-Hermitian skin effect~\cite{rmp-93-015005,prl-116-133903,prl-125-126402,nphys-16-761}, topological energy transfer~\cite{nature-537-80,ncomm-8-14154}, and asymmetric
mode switching~\cite{nature-537-76,prx-8-021066,prl-124-153903,prl-125-187403,nature-562-86,prl-86-787,optica-6-190,
prl-126-170506}, providing new opportunities for such a wide range of fields as synthetic photonics and topological physics~\cite{prl-127-250402,prl-127-196801,arxiv-2007-05960}.
However, due to the accumulation of dissipations in topological operations, as far as we know, topological amplifier which works as
a key element in practical application has remained a challenge as topological EP devices till now.

In this work, we propose how to achieve synthetic $\mathcal{PT}$ symmetry and topological amplifier in optomechanics~\cite{rmp-86-1391,pt-65-29,ap-525-215}, without the need of any active medium. We find that the optomechanical Stokes processes can be harnessed to compensate the optical losses and thus realize $\mathcal{PT}$ symmetry in such a passive system~\cite{prl-104-083901,nphoton-12-479} without complexities, such as fabricating gain materials in active systems~\cite{science-363-42,prl-118-093002,ncomm-7-13662}. As another merit, topological optical amplifications can be realized here by simply tuning the optical modes rather than steering the acoustic modes~\cite{nature-537-80} or designing materials with modulated structures~\cite{ncomm-8-14154,nature-537-76,prx-8-021066,prl-124-153903,prl-125-187403,nature-562-86,optica-6-190,prl-86-787,
prl-126-170506}. Our work confirms that optomechanical systems can serve as a powerful tool to observe and utilize various topological EP effects.

In comparison with the previous works for the EP, our scheme owns significant differences as follows. First of all,
in our work there exists an optical gain with a tunable center frequency as the special character of cavity optomechanics via a tunable frequency of the pump field.
It is beyond the traditional gain processes, especially for the one in cavity optomechanics~\cite{nphoton-12-479}, where the center
frequency of gain cannot be tuned for the certain frequency of the pump field. Secondly, different from the previous works for topological energy
transfer in waveguides~\cite{ncomm-8-14154,nature-537-76,prx-8-021066,
prl-124-153903,prl-125-187403,nature-562-86,optica-6-190,prl-86-787,prl-126-170506} or optomechanical
phonon modes~\cite{nature-537-80}, which are limited by the accumulation of dissipations, our work
illustrates that the dissipation accumulation can be overcome by employing the time-dependent gain with a tunable center frequency from the Stokes processes.
Thirdly, contrary to the path-dependent topological dynamics with waveguides, where these fabricated optical systems lose their tunability, our optical system is time-dependent, and thus feasible to simulate the topological dynamics with different trajectories and topological properties.
Finally, different from the previous work for energy transfer~\cite{nature-537-80}, where the effective two-level structure of phonon modes limits its application on the topological amplifier, our system benefits from the configuration of micro-toroidal resonators. This configuration leads to forward and backward transmissions taking different physical dynamics, and enables our system to work as a topological amplifier.

\section{Synthetic $\mathcal{PT}$ symmetric optomechanics}
\begin{figure}[h]
\centering \includegraphics[width=8cm]{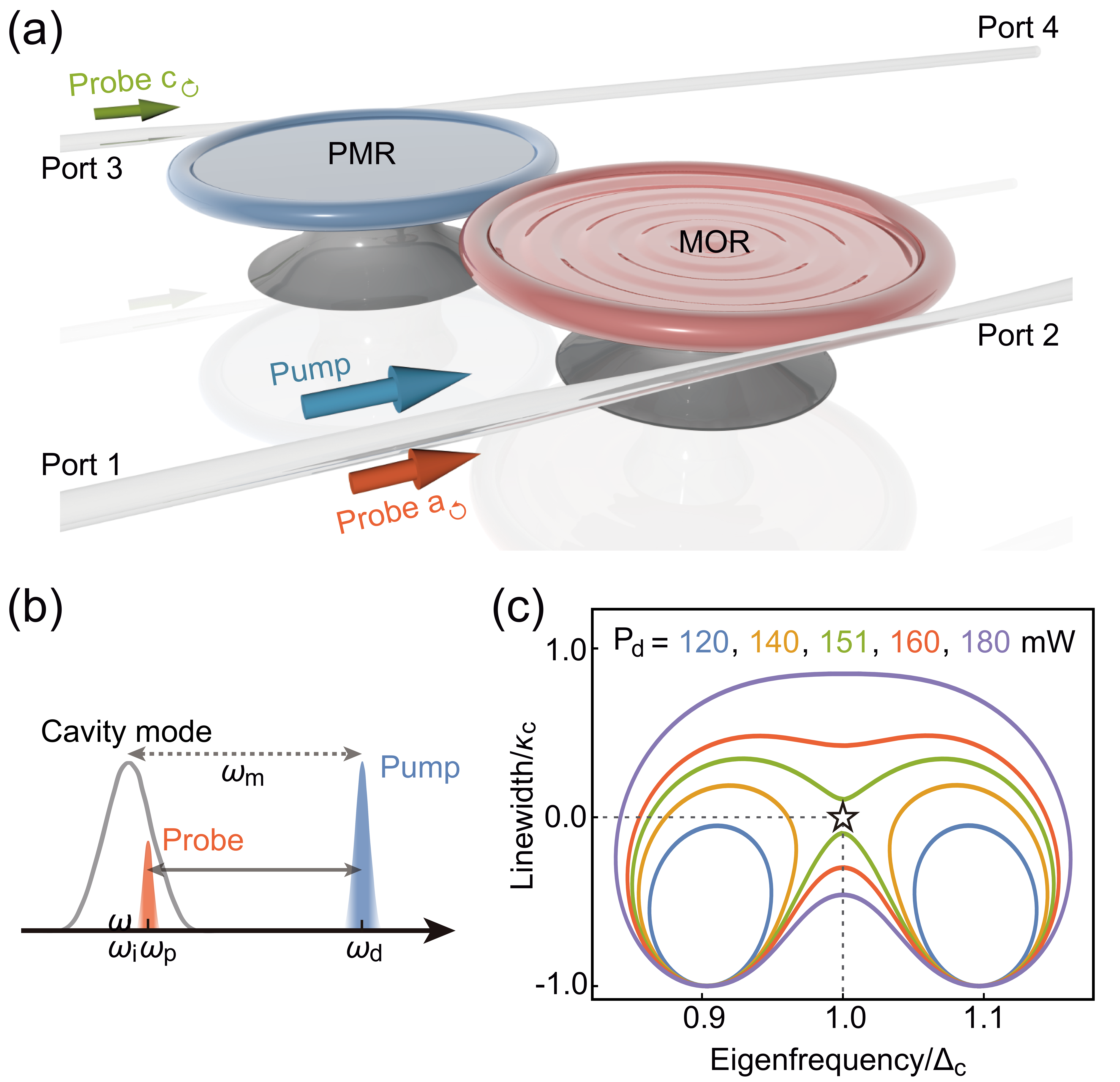}
\caption{Synthetic $\mathcal{PT}$
	symmetry in a passive optomechanical system without gain materials. (a) Two resonators, denoted by MOR and PMR, are evanescently coupled with each other and also coupled with optical fibers. A blue-detuned pump is input at the port $1$ and the probe field enters from the ports $1$ and $3$. (b) In the propagation direction of the pump, the low-frequency probe acquires
    an effective gain from the pump via the Stokes process.
	(c) The eigenfrequency $\mathrm{Re}\left[\omega\right]$ and its linewidth $\mathrm{Im}\left[\omega\right]$
	are the functions of the detuning $\Omega$ for different values of the
	pump power $P_\mathrm{d}$.}  %  Closed loops for topological operations on the Riemann surface of eigenenegies $\mathrm{Re}\left[\omega\right]$ in parameter spaces for the effective detuning $\Delta_\mathrm{eff}$ and decay rate $\kappa_\mathrm{eff}$.
\label{model}
\end{figure}

We start by considering a passive optomechanical system as shown in Fig.~\ref{model}(a), where a micro-toroidal optomechanical resonator (MOR) evanescently couples to a passive micro-toroid resonator (PMR)~\cite{prl-104-083901,nphoton-12-479}. This system can be described in the simplest level by the Hamiltonian
\begin{equation}
    H=\frac{p^{2}}{2m}+\frac{1}{2}m\omega_\mathrm{m}^{2}q^{2}-\chi
    		qa_\mathrm{\circlearrowleft}^{\dag}a_\mathrm{\circlearrowleft}+H_\mathrm{c},
     \label{ham0}
\end{equation}
where $q$ and $p$ are position and momentum operators of the vibrational mode, respectively. The vibrational mode takes an effective mass $m$ and an eigenfrequency $\omega_\mathrm{m}$.
The optical mode $a_\mathrm{\circlearrowleft}$ in MOR couples to the vibrational mode
via a radiation pressure coupling $\chi$. Optical mode $a_\mathrm{\circlearrowleft}$ is counter clockwise at frequency $\omega_\mathrm{a}$, which is
driven (detected) by a pump (probe) field with frequency $\omega_\mathrm{d}$ ($\omega
_\mathrm{p}$) and amplitude $\sqrt{2\kappa_\mathrm{a}}s_\mathrm{d}$
($\sqrt{2\kappa_\mathrm{a}}s_{a_\mathrm{\circlearrowleft}}$) from input port $1$,
while the optical mode $c_\mathrm{\circlearrowright}$ of PMR in clockwise at frequency $%
\omega_\mathrm{c}$ is only detected by a probe field with frequency $\omega_\mathrm{p}$
and amplitude $\sqrt{2\kappa_\mathrm{c}}s_{c_\mathrm{\circlearrowright}}$ from input port $3$. In the
rotating frame at frequency $\omega_\mathrm{d}$, the Hamiltonian for the cavities is given by
\begin{equation}
\begin{array}{ccl}
H_c/\hbar&=&\Delta
		_\mathrm{a}a_\mathrm{\circlearrowleft}^{\dag}a_\mathrm{\circlearrowleft}
		+\Delta
		_\mathrm{c}c_\mathrm{\circlearrowright}^{\dag}c_\mathrm{\circlearrowright}+
g(a_\mathrm{\circlearrowleft}^{\dag}c_\mathrm{\circlearrowright
		}+\mathrm{H.c.})\\
&&+i\sqrt{2\kappa_\mathrm{a}}s_\mathrm{d}(a_\mathrm{\circlearrowleft}^{\dag}-a_\mathrm{\circlearrowleft})\\
&&+i\sqrt{2\kappa_\mathrm{a}}s_{a_\mathrm{\circlearrowleft}}(a_\mathrm{\circlearrowleft}^{\dag}e^{-i\Omega t}-\mathrm{H.c.})\\
&&+i\sqrt{2\kappa_\mathrm{c}}s_{c_\mathrm{\circlearrowright}}(c_\mathrm{\circlearrowright}^{\dag}e^{-i\Omega t}-\mathrm{H.c.}),
\end{array}
\end{equation}
where $a_\mathrm{\circlearrowleft}$ ($a_\mathrm{\circlearrowleft
}^{\dag}$) and $c_\mathrm{\circlearrowright}$ ($c_\mathrm{\circlearrowright}^{\dag}$)
are the annihilation (creation) operators of MOR and PMR, respectively. $\Omega=\omega
_\mathrm{p}-\omega_\mathrm{d}$ ($\Delta_\mathrm{i=a,c}=\omega_\mathrm{i}-\omega_\mathrm{d}$)
is the detuning between the fixed probe field (cavity modes) and the tunable pump field. $g$ is
the evanescent coupling between MOR and PMR. $s_\mathrm{i}=%
\sqrt{P_\mathrm{i}/\hbar \omega_\mathrm{i}}$ is governed by power $P_\mathrm{i}$ for
$\mathrm{i}=\mathrm{p},a_\mathrm{\circlearrowleft},c_\mathrm{\circlearrowright}$, and $
\kappa_\mathrm{a (c)}$ is the decay rate for mode $a_\mathrm{\circlearrowleft}$ ($c_\mathrm{\circlearrowleft}$).

In the blue-sideband regime ($\Omega \simeq -\omega_\mathrm{m}$), by employing the
mean-value equations for Hamiltonian~(\ref{ham0}) and eliminating the vibrational mode, we obtain the effective mean-value equations for optical modes in the frequency domain as
~\cite{SM,walls2007quantum}
\begin{equation}
\begin{array}{ccl}
-i\Omega a_\mathrm{\circlearrowleft}&=&- (\kappa_\mathrm{eff}+i\Delta
_\mathrm{eff})a_\mathrm{\circlearrowleft}-igc_\mathrm{\circlearrowright}+\sqrt{2\kappa_\mathrm{a}}
s_{a_\mathrm{\circlearrowleft}},\\
-i\Omega c_\mathrm{\circlearrowright}&=&- (\kappa_\mathrm{c}+i\Delta
_\mathrm{c})c_\mathrm{\circlearrowright}-iga_\mathrm{\circlearrowleft}+\sqrt{2\kappa_\mathrm{c}}s_{c_\mathrm{\circlearrowright}},%
\end{array}
\label{Eq21}
\end{equation}
where the effective detuning and gain are, respectively,
\begin{equation}
\Delta_\mathrm{eff}=\beta \omega_\mathrm{m}\sin {\theta /}|\Omega_\mathrm{m}|+\Delta_\mathrm{a}-\chi q_\mathrm{s}/\hbar
\end{equation} and
\begin{equation}
\kappa_\mathrm{eff}=\kappa_\mathrm{a}-\beta \omega_\mathrm{m}\cos {\theta /|\Omega_\mathrm{m}|}
\end{equation} with $
\Omega_\mathrm{m}=\gamma_\mathrm{m}/2-i (\Omega +\omega
_\mathrm{m})$,
$\beta=\chi q_\mathrm{s}/ (2\hbar )$, $e^{i\theta}=\Omega_\mathrm{m}/|\Omega_\mathrm{m}|$, and $q_\mathrm{s}$ being
the steady-state position~\cite{SM}. The effective detuning $\Delta_\mathrm{eff}$ can be adjusted by mean photon number via the optomechanical interaction. As sketched in Fig.\,1(b),
the Stokes photons are created at frequency $\omega_\mathrm{p}\simeq\omega_\mathrm{d}-\omega_\mathrm{m}$ by emitting phonons at frequency $\omega_\mathrm{m}$,
resulting in an effective gain $\kappa_\mathrm{eff}$ for the probe field. This provides a natural way to reach the gain-loss balance or $\mathcal{PT}$ symmetry, which is
fundamentally different from the previous works using active materials~\cite{science-346-975,science-346-972,prl-113-053604}, tunable dissipation in passive cavities~\cite{nphoton-12-479}, and modulated structures~\cite{ncomm-8-14154,nature-537-76,prx-8-021066,prl-124-153903,prl-125-187403,nature-562-86,optica-6-190,prl-86-787,
prl-126-170506}.

In the adiabatic limit, the non-Hermitian Hamiltonian of optical modes $a_\mathrm{\circlearrowleft}$ and $c_\mathrm{\circlearrowright}$ for Eq.~(\ref{Eq21})
can be written in a time-dependent manner~\cite{prl-118-093002,SM}
\begin{equation}
\begin{array}{ccl}
	H_\mathrm{eff}(t)&=&\left(
	\begin{array}{cc}
	\Delta_\mathrm{eff}(t)-i\kappa_\mathrm{eff}(t)&g \\
	g&\Delta_\mathrm{c}-i\kappa_\mathrm{c}
	\end{array}
	\right)%
\end{array},\label{Eq23}
\end{equation}
which has the eigenmodes
\begin{equation}
	\left\vert \psi_\mathrm{\pm}(t)\right\rangle=(-i\lambda(t)\pm\sqrt{1-\lambda(t)
		^{2}})\left\vert a_\mathrm{\circlearrowleft}\right\rangle+\left\vert
	c_\mathrm{\circlearrowright}\right\rangle,\label{eigenmode}
\end{equation}
and the eigenvalues
\begin{equation}
	\omega=\omega_\mathrm{\pm}=V(t)/2\pm g\sqrt{1-\lambda(t) ^{2}}.\label{eigenvalues}
\end{equation}
Here
\begin{equation}
V(t)=\Delta_\mathrm{eff}+\Delta_\mathrm{c}-i(\kappa_\mathrm{eff}+\kappa_\mathrm{c}),
\end{equation}
and
\begin{equation}
\lambda(t)=\left[\kappa_\mathrm{eff}(t)-\kappa_\mathrm{c}+i (\Delta_\mathrm{eff}(t)-\Delta
_\mathrm{c})\right]/2g.
\end{equation}
$\left\vert a_\mathrm{\circlearrowleft}\right\rangle$ and
$\left\vert c_\mathrm{\circlearrowright}\right\rangle$ denote optical modes
$a_\mathrm{\circlearrowleft}$ and $c_\mathrm{\circlearrowright}$, respectively.

\begin{figure*}[h]
	\centering\includegraphics[width=16cm]{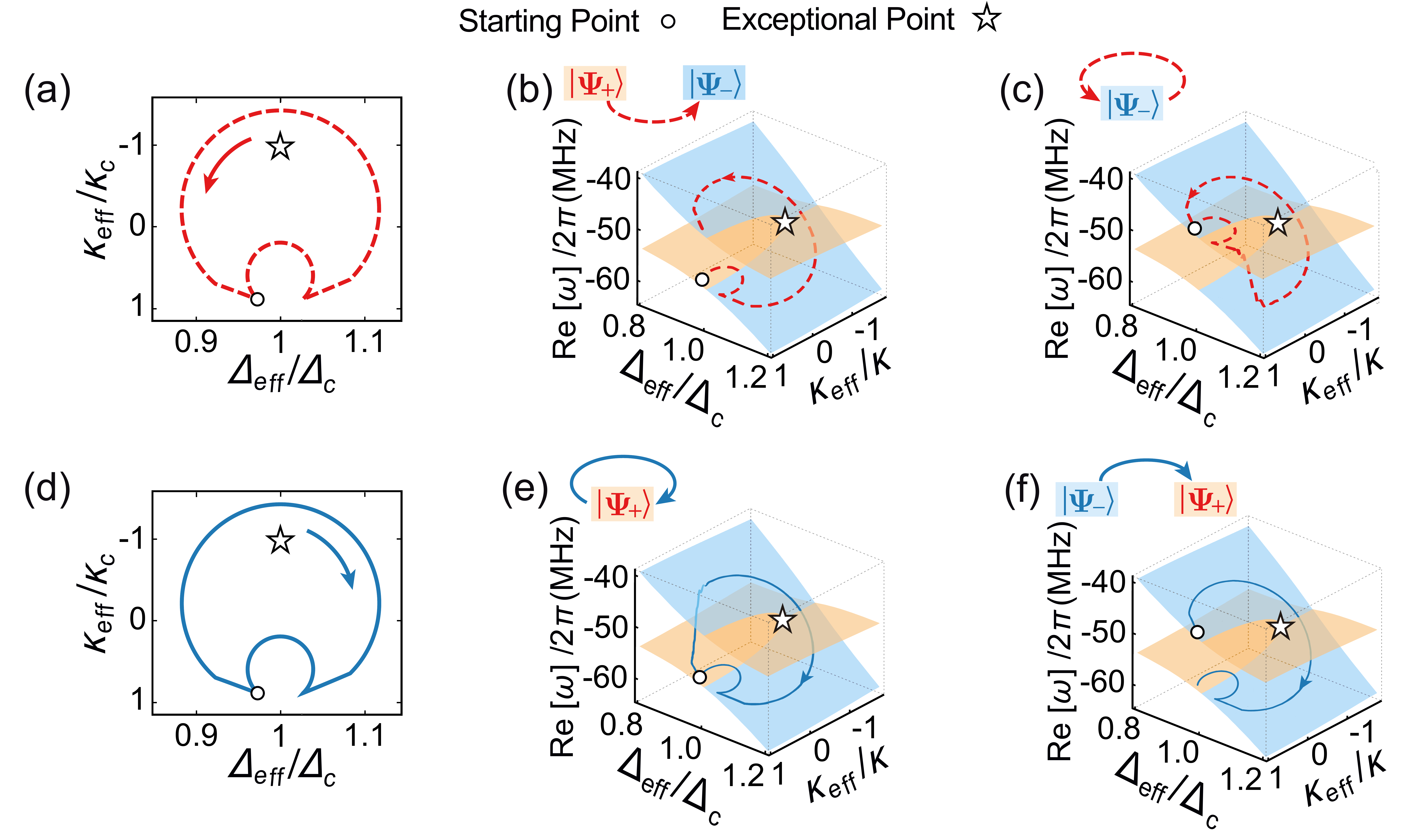}
	\caption{Energy dynamics $\mathrm{Re}\left[\omega\right]$ and their trajectories with chirality follow the loops of topological operations
		with EP. (a,d) Loops of the topological operation, where
     		the topological trajectories start from the initial eigenmodes plotted in the following panels. (b,e) Initial eigenmode
		$|\protect\psi_\mathrm{+}(t=0)\rangle$ and (c,f) initial eigenmode $|\protect\psi_\mathrm{-}(t=0)\rangle$
        	in the parameter spaces of the eigenenergy $\mathrm{Re}\left[\omega\right]$,
		effective detuning $\Delta_\mathrm{eff}$ and decay rate $\protect\kappa_\mathrm{eff}$.
		Arcs with arrows represent the topological operation directions in CCW (dashed red) and CW (solid blue).
		Here eigenenergies for eigenmodes $|\protect\psi_\mathrm{+}\rangle$ and $|\protect\psi_\mathrm{-}\rangle$
		are plotted in orange and blue. Parameters for
		simulations are $T=0.01~\mathrm{ms}$, $\protect\omega_\mathrm{m}/2\protect\pi=51.8~\mathrm{MHz}$,
		$\protect\gamma_\mathrm{m}/2\protect\pi=41~\mathrm{kHz}$, $m=20~\mathrm{ng}$, $\Delta_\mathrm{c}=-\protect\omega_\mathrm{m}$,
		$g/2\protect\pi=5~\mathrm{MHz}$, $\protect\kappa /2\protect\pi=5~\mathrm{MHz}$,
		$\protect\kappa_\mathrm{c}=\protect\kappa_\mathrm{a}=\protect\kappa $, $\protect\lambda=390~\mathrm{nm}$,
		$\protect\chi /2\protect\pi=12\times10^{18}\hbar$~\cite{science-330-1520,nphys-10-394,nphoton-12-479,prl-104-083901,nphoton-12-479}.}
	\label{char}
\end{figure*}

The topological features of Hamiltonian~(\ref{Eq23}) can be identified from the complex eigenvalues~(\ref{eigenvalues}) versus detuning $\Omega$ with
different pump power $P_\mathrm{d}$ as plotted in Fig.~\ref{model}(c). Two independent orange circles ($P_\mathrm{d}=140~\mathrm{mW}$) % for eigenmodes
are gradually melting into a big green circle ($P_\mathrm{d}=151~\mathrm{mW}$) with the increase of
the pump power $P_\mathrm{d}$. It is the larger pump power $P_\mathrm{d}$ that provides a larger effective gain and
ensures the EP to be enclosed in closed circles. In addition, topological features can also be identified from the Riemann surface in Fig.~\ref{char}.

\section{Topological engineering around EPs}

To describe the topological dynamics around the EP,  we need to acquire the effective scattering matrix of Hamiltonian~(\ref{Eq23}) at first.

We assume an evolution trajectory consisting of N short sections, where the scattering matrix regarding the section $k$ is given by
\begin{equation}
\begin{array}{l}
	U^\mathrm{eff}_{k}=exp[-i(\delta\Delta/2-i\delta\kappa/2)T_\mathrm{0}]\\
*
    \left(
	\begin{array}{cc}
	cos\frac{\lambda T_\mathrm{0}}{2}-\eta sin\frac{\lambda T_\mathrm{0}}{2}& -\frac{2ig}{\lambda}sin\frac{\lambda T_\mathrm{0}}{2} \\
-\frac{2ig}{\lambda}sin\frac{\lambda T_\mathrm{0}}{2}&cos\frac{\lambda T_\mathrm{0}}{2}+\eta sin\frac{\lambda T_\mathrm{0}}{2}
	\end{array}
	\right)%
\end{array}
\end{equation}
with
\begin{equation}
\left\{
\begin{array}{ccl}
\lambda&=&\sqrt{4g^\mathrm{2}+(\delta\Delta-i\delta\kappa)^2},\\
\eta&=&\frac{i\delta\Delta+\delta\kappa}{\lambda},\\
\delta\Delta&=&\Delta_\mathrm{eff}[(k-1)T_\mathrm{0}]-\Delta_\mathrm{c},\\
\delta\kappa&=&\kappa_\mathrm{eff}[(k-1)T_\mathrm{0}]-\kappa_\mathrm{c},
\end{array}\right.
\end{equation}
where $T_\mathrm{0}$ is the evolution time for each section.

The corresponding total scattering matrix can be written as
\begin{equation}
U=\Pi_{k=1}^{N} U^\mathrm{eff}_{k}\label{matrix},
\end{equation}
which shows that the amplification of the transmission field is determined by the time-dependent net gain $\delta\kappa$ via
the Stokes processes, and the time-dependent evolution mode $|\psi(t)\rangle$ can be expressed as
\begin{equation}
|\psi(t)\rangle=U|\psi(t=0)\rangle,
\end{equation} with an initial mode $|\psi(t=0)\rangle$ for the time $t=0$.

Next, to illustrate the NATs in the topological dynamics of our system, we simulate the evolution trajectories for
topological operations in counter-clockwise (CCW) and clockwise (CW) following the scattering matrix~(\ref{matrix}) with $\mathrm{N}=4000$ in Fig.~\ref{char}.

Fig.~\ref{char} indicates some trajectories between two eigenenergy surfaces in the parameter space as shown in Fig.~\ref{char}(c,e). These trajectories from the unsteady eigenmode
$|\psi_\mathrm{+}\rangle$ to the steady one $|\psi_\mathrm{-}\rangle$ are regarding NATs. The NATs will appear when
the topological operation times is longer than the coherence times of the optical modes~\cite{pra-92-052124}.
Therefore, NATs enable topology-dependent energy transfers and own chiral properties in the dynamical encircling of the EP in the parameter space.

More specifically, our system is dominated by loss ($|\kappa_{\mathrm{eff}}|<\kappa$).
When the high energy surface is in $|\psi_\mathrm{+}\rangle$ ($\Delta_\mathrm{eff}>\Delta_\mathrm{c}$),
the high energy eigenmode $|\psi_\mathrm{+}\rangle$, following the topological operations in CCW [see Fig.~\ref{char}(a)], will decay to its
steady eigenmode $|\psi_\mathrm{-}\rangle$ as in Fig.~\ref{char}(c). That is a traditional transition regarding dissipations. In contrast,
when high energy surface is in $|\psi_\mathrm{-}\rangle$ ($\Delta_\mathrm{eff}<\Delta_\mathrm{c}$),
the low energy eigenmode $|\psi_\mathrm{+}\rangle$, following the topological operations in CW [see Fig.~\ref{char}(d)], will decay to the high
energy eigenmode $|\psi_\mathrm{-}\rangle$ as in Fig.~\ref{char}(e). That works as the counterintuitive NAT
from the low energy surface to the high one. This counterintuitive transition, resulting from detuning ($\Delta_\mathrm{eff}-\Delta_\mathrm{c}<0$),
ensures the steady eigenmode with a higher eigenenergy as in Eq.~(\ref{eigenvalues}).

The eigenenergy surface in Fig.~\ref{char} also indicates that the EP of synthetic $\mathcal{PT}$ symmetric optomechanics
can be induced by the effective gain from Stokes processes
via radiation pressure coupling. This effective gain takes tunable frequency for the radiation pressure coupling.
That is different from the traditional gain methods based on rare-earth-doped gain media~\cite{prl-114-253601,prl-117-110802,science-346-972,science-346-975,prl-113-053604} and
stimulated Brillouin processes~\cite{nature-576-65}, where the effective gain is limited by the frequencies of optical modes and gain materials. Moreover, in comparison to the conventional ideas induced by the coupling strength ~\cite{nphys-10-394,nphoton-8-524} and the loss ~\cite{prl-103-093902,nmat-12-108,science-346-328},
our system provides an alternative way to observe $\mathcal{PT}$-symmetric breaking with the effective gain.

\begin{figure*}[tbph]
	\centering\includegraphics[width=16cm]{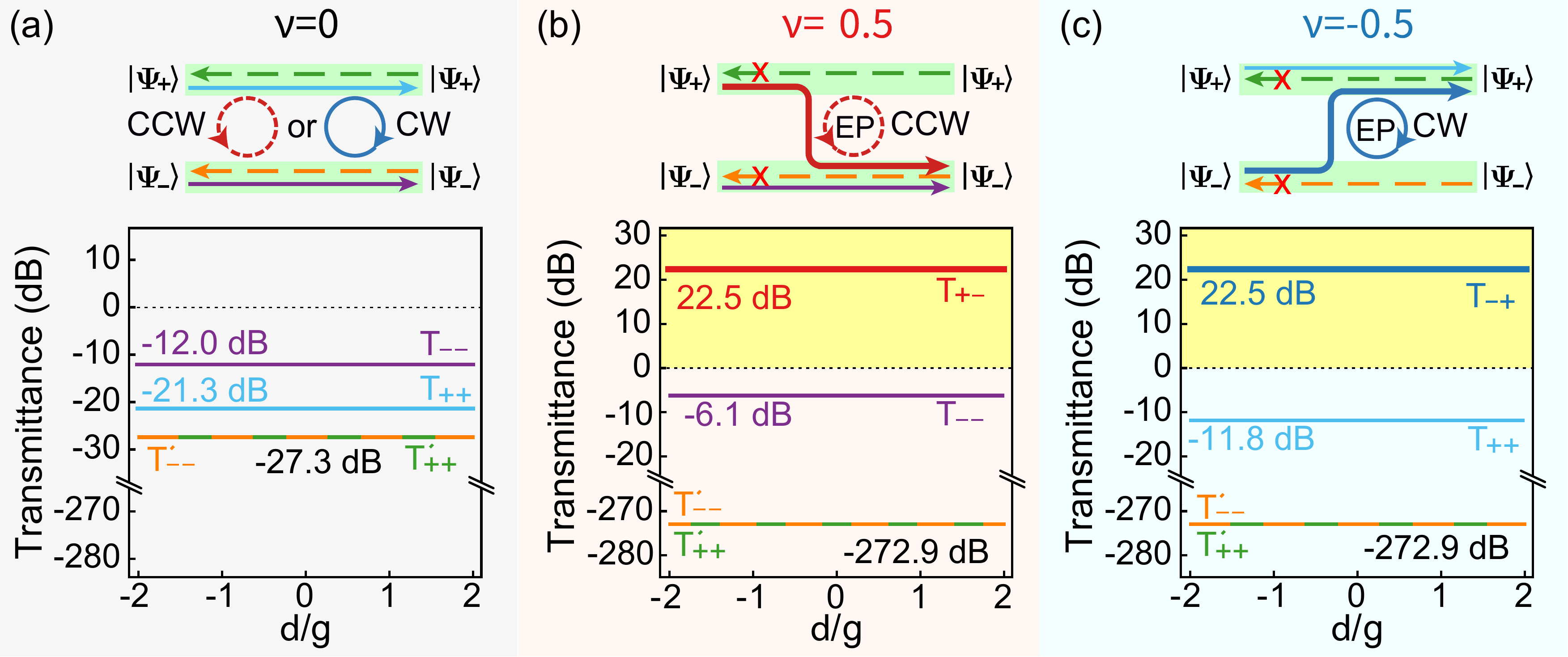}
\caption{Transmission spectra $T_\mathrm{mn}$ from the input eigenmode of forward transmissions $|\psi_\mathrm{m}(t=0)\rangle$
		to the output eigenmode $|\psi_\mathrm{n}(t=T)\rangle$ with $m,n=\pm$. Different topological dynamics is sorted via invariant vorticity $\nu$. (a) Reciprocal transmission spectra $T_\mathrm{mn}$ for trival topological dynamics in CCW and CW ($\nu=0$) without the EP for the topological operation time of $T=0.1~\mathrm{\mu s}$ with $\Delta_{\mathrm{eff}}/\Delta_{\mathrm{c}}\in\{0.9,0.92\}$ and $\kappa_{\mathrm{eff}}/\kappa_{\mathrm{c}}\in\{0.15,0.3\}$~\cite{SM}; (b,c) Non-reciprocal transmission spectra $T_\mathrm{mn}$ for non-trival topological dynamics encircling the EP in (b) CCW ($\nu=0.5$) and (c) CW ($\nu=-0.5$) with $T=1~\mathrm{\mu s}$, $\Delta_{\mathrm{eff}}/\Delta_{\mathrm{c}}\in\{0.9,1.1\}$ and $\kappa_{\mathrm{eff}}/\kappa_{\mathrm{c}}\in\{-1.2,0.3\}$~\cite{SM}. The transmission spectra are characterized with
        $T_\mathrm{mn}=\langle\psi_\mathrm{m}(t=0)|\psi_\mathrm{n}(t=T)\rangle$. Detuning $\delta$ is the deviation frequency between the real probe pulses
		and the ideal ones at frequency $\omega_\mathrm{p}$, and vorticity number $\nu$ is a topological invariant~\cite{prl-120-146402,prl-123-066405,SM}.
		The effective parameters satisfy the approximate conditions of the theoretical derivations in supplemental materials~\cite{SM}.
		Other parameters take the same values as in Fig. \ref{char}.}
	\label{spectra}
\end{figure*}

%\section{Topological number: Vorticity $\nu$ for topological invariant}
To illustrate the non-trivial topological properties of our system, we have to introduce the topological number.
Here we can define the topological number as vorticity $\nu$ in Ref.~\cite{prl-120-146402}.

According to Ref.~\cite{prl-120-146402}, to show the topological invariant of the topological operations, we employ the invariant vorticity $\nu$
for eigenenergies $\omega_{\pm }$ in the complex-energy plane as%
\begin{equation*}
\nu(\alpha )=-\dfrac{1}{2\pi }\oint\limits_{\alpha }\triangledown_{k}\arg
[\omega_{+}(k)-\omega_{-}(k)]dk,
\end{equation*}
where $\alpha$ is a closed loop in the complex-energy plane, and $k=\Omega $ ($k=P_{d}$) is for the fixed pump field $P_{d}$ (detuning $\Omega$). This equation shows that the EP is (isn't) enclosed in the loops of the complex-energy plane for $\nu=\pm0.5$ ($\nu=0$). Then we can obtain the red dash curves and the solid blue one in
Fig.~\ref{char}, corresponding to $\nu=0.5$ and $\nu=-0.5$, respectively. In addition, it is worthy to point out that the topological feature can
also be observed from the linewidth and eigenenergy of the eigenmodes in Fig. 1(c), where two independent orange circles ($P_{d}=140$ mW) for eigenmodes will melt into a big green curve ($P_{d}=151$ mW) by increasing the pump power.

\section{Numerical simulation and discussion}
In this section, we demonstrate our synthetic $\mathcal{PT}$ symmetric system to work as a topological amplifier with numerical simulations.

\subsection{Topological energy transfer}
To illustrate topological energy transfer and its chiral properties, we plot the trajectories for the loops of topological operations
enclosing an EP in CW and CCW as in Fig.~\ref{char}.

When the EP is enclosed in the loops of topological operations, we assume the topological operation
time $T$ to be long enough in accomplishment of a single NAT. In this case, topological trajectories for different
initial eigenmodes depend on the topological operation direction and own chirality. For example, only the
initial mode $|\psi_\mathrm{+}(t=0)\rangle$ [$|\psi_\mathrm{-}(t=0)\rangle$] evolving along the
Riemann surface in CCW (CW) can be transferred to $|\psi_\mathrm{-}(t=T)\rangle\simeq|\psi_\mathrm{+}(t=0)\rangle$ as in Fig.~\ref{char}(b)
[($|\psi_\mathrm{+}(t=T)\rangle\simeq|\psi_\mathrm{-}(t=0)\rangle$ as in Fig.~\ref{char}(f)].
Otherwise, the optical mode will return to its initial mode $|\psi_\mathrm{+}(t=T)\rangle\simeq|\psi_\mathrm{+}(t=0)\rangle$
as in Fig.~\ref{char}(c) [$|\psi_\mathrm{-}(t=T)\rangle\simeq|\psi_\mathrm{-}(t=0)\rangle$
as in Fig.~\ref{char}(e)] since the NAT process blocks the eigenmodes swapping along the Riemann surface.
As a result, topological energy transfer between two optical modes can be achieved, taking the feature of chirality. The above topological energy transfer and chirality can also be
understood from the combination of the unsteady and steady eigenmodes, similar to the
results characterized experimentally~\cite{nature-537-76,nature-562-86,prx-8-021066,prl-124-153903,prl-125-187403}.
Due to this reason, this chirality can be switched off by further increasing
the topological operation time $T$~\cite{SM}, as predicted in Ref.~\cite{prl-118-093002}.

%{\color{blue}
%\section{Time-dependent topological optomechanical amplification}

\subsection{Topological amplification}
The topological energy transfer mentioned above inspires us to achieve time-dependent topological optomechanical
amplification for optical probe pulses with tunable topological properties by designing loops enclosing EP with large enough gain in the parameter space.
Therefore, we will employ probe pulses to illustrate the time-dependent topological optomechanical amplification as follows.
%}

To quantify the performance of time-dependent topological optomechanical amplification under the influence of
the probe field frequency, we plot transmission spectra for probe pulses in Fig.~\ref{spectra}
with the scattering matrix~(\ref{matrix}) and experimentally achievable parameters~\cite
{science-330-1520,nphys-10-394,nphoton-12-479,prl-104-083901},
which ensures the largest effective gain for simulation to be $\kappa_\mathrm{eff}/2\pi\simeq-7.1~\mathrm{MHz}$~\cite{SM}
with the pump power $P_\mathrm{d}=180~\mathrm{mW}$. Then the features of tunability, chirality, and topology can be illustrated
from the transmissions in the topological dynamics with and without the EP for different values of vorticity number
$\nu$~\cite{prl-120-146402,prl-123-066405,SM} as elucidated below. In the following, we assume the input probe pulses at
ports 1(2) and 3(4) are in forward transmission eigenmodes $|\psi_\mathrm{\pm}(t=0)\rangle$ of the system~\cite{nature-537-76,nature-562-86,prx-8-021066,prl-124-153903,prl-125-187403}.

When EP is not encircled by the trajectories of topological operations ($\nu=0$), the topological operations can be
accomplished within the coherence time in the adiabatic limit, as mentioned by the
traditional adiabatic theory~\cite{walls2007quantum}. This means that our system can evolve along the eigenenergy surface $|\psi_\mathrm{\pm}\rangle$ and return
to its initial modes $|\psi_\mathrm{\pm}(t=T)\rangle=|\psi_\mathrm{\pm}(t=0)\rangle$
without the NAT processes (see Fig. S5 in~\cite{SM}).
In other words, no chirality exists in the transmissions as in Fig.~\ref{spectra}(a), where
all the transmission spectra for topological operations in CCW and CW share the same
value $T_\mathrm{mn}\leq-12~\mathrm{dB}$ since the loss dominates the system in this case.

On the other hand, when EP is enclosed by the trajectories of topological operations ($\nu=\pm0.5$),
the chiral amplification of the energy can be found from the spectra difference between
the forward transmissions $T_{+-}$ ($T_{-+}$) and $T_{--}$ ($T_{++}$),
which are input from ports \{1,3\} and output at ports \{2,4\}.
Specifically, only the transmission $T_{+-}$ ($T_{-+}$) from the initial eigenmode
$|\psi_\mathrm{+}(t=0)\rangle$ ($|\psi_\mathrm{-}(t=0)\rangle$) to the final eigenmode
$|\psi_\mathrm{-}(t=T)\rangle$ ($|\psi_\mathrm{+}(t=T)\rangle$) in CCW (CW) can be
amplified, as indicated by the solid lines in Fig.~\ref{spectra}(b,c). Otherwise, the
transmission $T_{--}$ ($T_{++}$) from the initial eigenmode
$|\psi_\mathrm{-}(t=0)\rangle$ ($|\psi_\mathrm{+}(t=0)\rangle$) to the final
eigenmode $|\psi_\mathrm{-}(t=T)\rangle$ ($|\psi_\mathrm{+}(t=T)\rangle$) in CCW (CW)
will be suppressed by the NAT processes due to the combination of gain and loss, see the curves in Fig.~\ref{char}(c,e).
That is to say, the chirality regarding the dynamical enclosing of an EP implies that the final
eigenmodes are only determined by the directions of
the topological operations in CCW and CW, and the amplification (suppression) of the initial modes depends
on the directions of the topological operations, see red (dark blue) curves in Fig.~\ref{spectra}(b,c).
Nevertheless, these amplifications would be less than the ones in numerical simulations for the
limitation of saturation of Stokes processes~\cite{nphoton-12-479}, while the ratios of amplification of optical modes in output probe pulses
will keep in a constant, since the output probe pulses are always the eigenmodes of the system.

According to the above discussion, our topological amplification ($\nu\neq0$) results from the accumulation of the gain and loss via NAT processes. The NAT processes lead to our topological amplification beyond the adiabatic condition. Therefore, the shortest time for our topological operation is the time to finish a single NAT, i.e., constrained by both the loss and the effective gain.

\subsection{Robustness of the topological amplification}
\begin{figure*}[tbph]
	\centering\includegraphics[width=16cm]{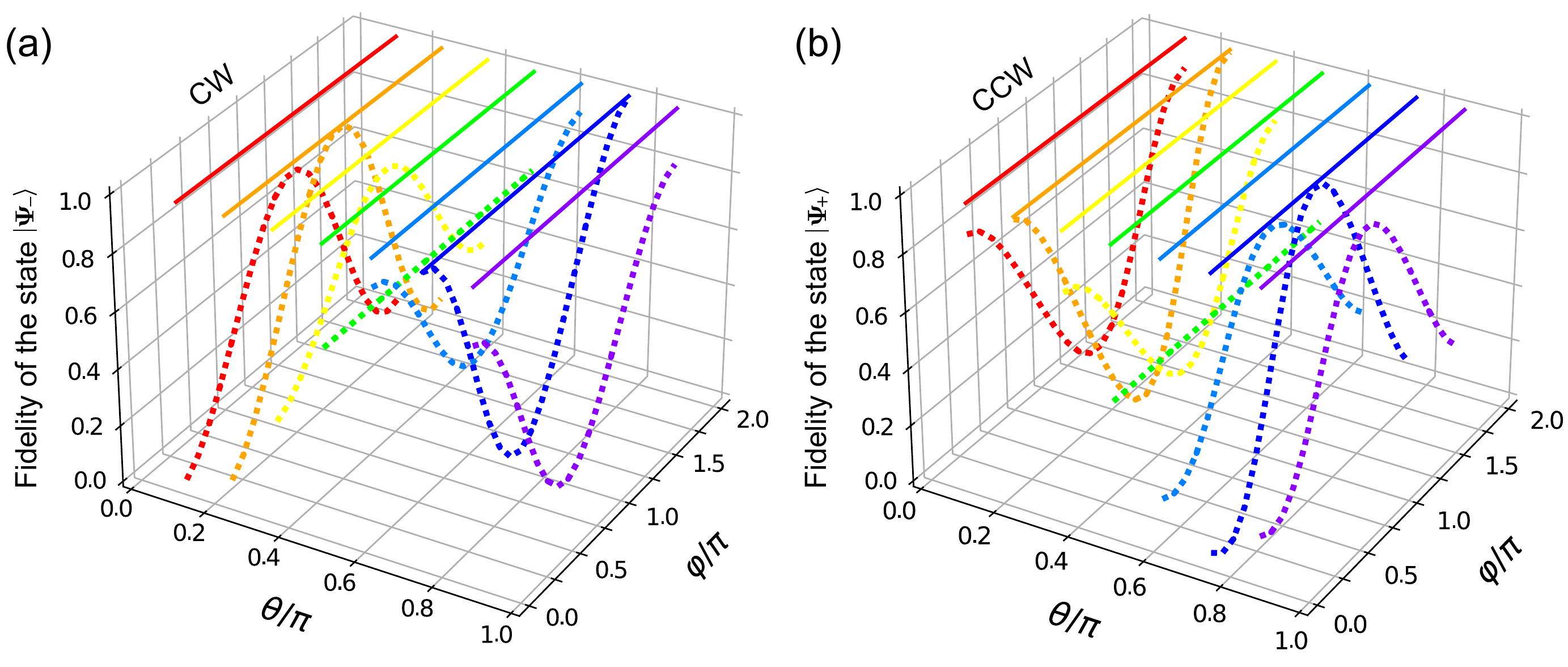}
\caption{The fidelity of different topological operations in (a) CCW and (b) CW with different initial states. Here solid lines represent topological amplification of the nontrival topological operations with an EP in (a) CCW ($\nu=0.5$) and (b) CW ($\nu=-0.5$), while the dotted curves illustrate the trival topological operations without and EP in CCW and CW ($\nu=0$). Parameters in the simulations are $\protect\omega_\mathrm{m}/2\protect\pi=51.8~\mathrm{MHz}$,
		$\protect\gamma_\mathrm{m}/2\protect\pi=41~\mathrm{kHz}$, $m=20~\mathrm{ng}$, $\Delta_\mathrm{c}=-\protect\omega_\mathrm{m}$,
		$g/2\protect\pi=5~\mathrm{MHz}$, $\protect\kappa /2\protect\pi=5~\mathrm{MHz}$,
		$\protect\kappa_\mathrm{c}=\protect\kappa_\mathrm{a}=\protect\kappa $, $\protect\lambda=390~\mathrm{nm}$,
		$\protect\chi /2\protect\pi=12\times10^{18}\hbar$, and the topological operation times are $T=10~\mathrm{\mu s}$ for (a,b) and $T=4.1~\mathrm{\mu s}$ for (c,d). Other parameters take the same values as in Fig. \ref{char}.}
	\label{fidelity}
\end{figure*}

Now we will demonstrate the robustness of the topological amplification with different initial modes by simulating the fidelity of the final mode $|\psi(t=T)\rangle=|\psi_{\pm}\rangle$ with different initial mode $|\psi(t=0)\rangle=\mathrm{cos}\theta|c_\mathrm{\circlearrowright}\rangle+e^{i\phi}\mathrm{sin}\theta|a_\mathrm{\circlearrowleft}\rangle$ in Fig.~\ref{fidelity}.

Fig.~\ref{fidelity} shows, when topological operations don't contain an EP, the initial mode is the same as the final one. It is due to the adiabatic condition. It can be identified from the oscillations of fidelities (see dotted curves). In particular, this result can be easily followed when the initial mode is in $|a_\mathrm{\circlearrowleft}\rangle$. In this case, the final mode is independent of the initial angle $\phi$ and takes a constant fidelity of $0.5$ (see dotted curves in green). On the other hand, when the topological operations contain an EP, the final modes must be the eigenmodes (see solid lines). We can select the eigenmodes via the evolution directions in CCW and CW (see Sec.4.2). These phenomena can also be understood as the robustness original from topological properties.

\subsection{Topological amplifier}
To show our system working as a topological amplifier, we have to demonstrate the above topological optomechanical amplification owning the feature of non-reciprocity. The backward transmissions, which are input from ports \{2,4\} and output at ports \{1,3\}, are denoted by the dashed lines in Fig.~\ref{spectra}(b,c) with the same input probe pulses as the forward transmission.
Since the backward optical modes $a_\mathrm{\circlearrowright}$ and $c_\mathrm{\circlearrowleft}$ decouple from
the forward ones $a_\mathrm{\circlearrowleft}$ and $c_\mathrm{\circlearrowright}$, the optical mode
$a_\mathrm{\circlearrowright}$ takes a certain decay rate $\kappa_\mathrm{a}$ and an effective detuning $\Delta_{\mathrm{eff}}$~\cite{SM}.
It implies that dissipations regarding the backward transmissions only depend on the topological operation time $T$, irrelevant to the input eigenmodes and the direction of topological operations. Therefore, the backward transmissions in CCW and CW share the same value of $T_{mn}=-272.9$ dB for the same topological operation time, see the dashed lines in Fig.~\ref{spectra}(b,d).
Here the transmission values of $T_{mn}$ are the ones for output pulses projected on the eigenmodes when the
topological operations are finished. As probe pulses suffer from long time dissipation, the output probe pulses would be
too small to be detected, and the non-reciprocal transmissions can be realized in this way.

Moreover, these final output probe pulses in Fig.~\ref{spectra} are insensitive to the detuning $\delta$. It is due to the fact that the final output probe pulses must be in one of the eigenmodes of the system, determined by the initial parameters of the pump field and selected by the direction of the topological operation.

As a result, our proposed synthetic $\mathcal{PT}$ symmetric optomechanics working as the topological optomechanical amplifier is feasible using current laboratory technologies~\cite{prl-104-083901,nphoton-12-479}. This optomechanical amplifier is based on the effective gain from the pump field, which opens a new way to overcome the low efficiency
transmission due to the loss accumulation ~\cite{nature-537-76,nature-562-86,prx-8-021066,prl-124-153903,prl-125-187403}. Compared with Ref.~\cite{nphoton-12-479} involving a certain gain from the certain pump field, our scheme is based on a time-dependent pump field and the tunable gain enables the optical amplification with topological properties. Therefore, the accumulation of dissipations for time-dependent topological operations~\cite{nature-562-86} can be overcome in this way.
Moreover, our topological amplifier owns features of tunability, chirality, and topology, which
are unattainable from the conventional Hermitian Hamiltonian devices~\cite{nphoton-10-657,ncomm-9-1797,nphoton-14-700,prl-120-023601}.
In addition, the optical gain can be enhanced by increasing the time for topological operations
via the effective gain~\cite{SM}, and the chirality can be switched off by increasing the topological operation time. Since the
final eigenmodes in our system must be in a steady eigenmode $|\psi_\mathrm{-}(t=T)\rangle$ when all NAT
processes are finished (see Fig. S3 in~\cite{SM}), our system offers the possibility to observe the predicted results of a long time
topological dynamics around the EP~\cite{prl-118-093002}, which overcomes the limitation of the traditional optical gain medium~\cite{science-363-42}.

\section{Conclusion}
We have explored how to achieve synthetic $\mathcal{PT}$ symmetry in passive optomechanics~\cite{prl-104-083901,nphoton-12-479}.
In such a system, we have demonstrated that topological dynamics around the EP can be selected and manipulated by a tunable blue-detuned pump field via Stokes processes from radiation pressure coupling, resulting in a topological amplifier. The proposed way of creating the effective optical gain via the
blue-detuned pump field can be applied to diverse systems with similar processes, such as stimulated Brillouin scattering~\cite{nphys-11-275,ncommu-6-6193},
stimulated Raman scattering~\cite{GPAgarwal}, and coupled nano-mechanical resonator array~\cite{nphys-9-480,scia-2-e600236}. Also, the synthetic $\mathcal{PT}$
symmetric optomechanics provides a new platform to explore time-dependent non-Hermitian dynamics~\cite{prl-126-170506} and topological photonics, with applications ranging from optical communications to quantum optical engineering.

\section*{Acknowledgement}
J. Q. Zhang appreciates helpful discussions with Jin-Hui Wu, Xiaoming Cai and Jian Xu.
J. Q. Zhang and M. Feng are supported by Special Project for Research and Development in Key areas of Guangdong Province (Grant No. 2020B030300001), the National Key Research and
Development Program of China (No. 2017YFA0304503), and the National Natural Science Foundation of China (Grant Nos. U21A20434, 91636220 and 11835011).
Z. R. Gong is supported by Natural Science Foundation of Guang-dong Province(Grant No. 2019A1515011400).
S. L. Su is supported by the National Natural Science Foundation of China (Grant No 11804308)
H. L. Zhang and H. Jing are supported by the National Natural Science Foundation of China (Grants No. 11935006 and No. 11774086)
and the Science and Technology Innovation Program of Hunan Province (Grant No. 2020RC4047).

\end{document}